# Properties and Geoeffectiveness of Magnetic Clouds during Solar Cycles 23 and 24

N. Gopalswamy[1], S. Yashiro[1,2], H. Xie[1,2], S. Akiyama[1,2], and P. Mäkelä[1,2]

[1]Solar Physics Laboratory, NASA Goddard Space Flight Center, Greenbelt, MD 20771

[2]The Catholic University of America, Washington DC 20064

Corresponding author email: nat.gopalswamy@nasa.gov

**Key points**

- Properties of magnetic clouds in cycle 23 and 24 are significantly different
- Anomalous expansion of CMEs cause the magnetic dilution in clouds
- Reduced magnetic content and speed of clouds lead to low geoeffectiveness
- 

Abstract

We report on a study that compares the properties of magnetic clouds (MCs) during the first 73 months of solar cycles 23 and 24 in order to understand the weak geomagnetic activity in cycle 24. We find that the number of MCs did not decline in cycle 24, although the average sunspot number is known to have declined by ~40%. Despite the large number of MCs, their geoeffectiveness in cycle 24 was very low. The average $Dst$ index in the sheath and cloud portions in cycle 24 was -33 nT and -23 nT, compared to -66 nT and -55 nT, respectively in cycle 23. One of the key outcomes of this investigation is that the reduction in the strength of geomagnetic storms as measured by the $Dst$ index is a direct consequence of the reduction in the factor $VB_z$ (the product of the MC speed and the out-of-the-ecliptic component of the MC magnetic field). The reduction in MC-to-ambient total pressure in cycle 24 is compensated for by the reduction in the mean MC speed, resulting in the constancy of the dimensionless expansion rate at 1 AU. However, the MC size in cycle 24 was significantly smaller, which can be traced to the anomalous expansion of coronal mass ejections near the Sun reported by Gopalswamy et al. (2014a). One of the consequences of the anomalous expansion seems to be the larger heliocentric distance where the pressure balance between the CME flux ropes and the ambient medium occurs in cycle 24.

## 1. Introduction

The properties of white-light coronal mass ejections (CMEs) have been reported to be significantly different between solar cycles 23 and 24 (Gopalswamy et al. 2014a). While the sunspot number declined by about 40% from cycle 23, the daily rate of CMEs remained roughly the same. For a given CME speed, the observed angular width of CMEs in cycle 24 was significantly larger than in cycle 23. Similarly, the number of halo CMEs, which constitute an energetic population among CMEs, also did not decline in cycle 24 (Gopalswamy et al. 2015a). These observations were explained by an anomalous expansion of CMEs as supported by the reduction in the ambient total pressure measured at 1 AU. The drastic reduction in the frequency and magnitude of major geomagnetic storms were attributed to the anomalous expansion of CMEs in cycle 24. Gopalswamy et al. [2014a] also compared the magnitudes of the southward component of the interplanetary magnetic field that caused major geomagnetic storms in cycles 23 and 24 and found that the $B_z$ values were generally lower in cycle 24. However, the sample size was very small in that study because only about a dozen major storms occurred in cycle 24.

Instead of starting with geomagnetic storms, it is worthwhile comparing the interplanetary structures related to CMEs that cause geomagnetic storms. In particular, we would like to consider magnetic clouds (MCs), which are a subset of interplanetary CMEs (ICMEs) possessing well-defined magnetic properties [e.g., Burlaga et al. 1981]: smooth rotation of one of the magnetic field components, enhanced magnetic field strength, and low proton temperature or low plasma beta. Ever since the identification of the southward pointing magnetic field in MCs [Wilson, 1987], there have been a large number of investigations associating MCs to geomagnetic storms [Zhang and Burlaga, 1988; Tsurutani et al. 1988; Echer et al. 2005; Yermolaev et al. 2007; Zhang et al. 2007; Gonzalez et al. 2011]. We consider MCs alone because of their simple structure amenable to elegant description [e.g., Goldstein 1983; Burlaga 1988; Lepping et al. 1990; Démoulin and Dasso 2009]. We also consider both shockless and shock-driving MCs because the latter have additional location (the sheath ahead of MCs) that may possess southward magnetic field and hence cause the so-called sheath storms [e.g., Kamide et al. 1998; Echer et al. 2008; Gopalswamy 2008; Yermolaev et al. 2012]. Many authors have studied the occurrence and strength of storms at various phases of solar cycles [e.g., Le et al. 2013; Kilpua et al. 2015 and references therein]. Le et al. [2013] defined the solar maximum as a single instance coinciding with the first of the typical two peaks during the sunspot maximum phase. However, the solar maximum is actually an extended period about 3 years, so some of their conclusions regarding the occurrence of intense storms in the decay phase may actually correspond to the maximum phase. Kilpua et al. [2015] have shown that stronger storms tend to occur during the maximum phase of the solar cycle. This is understandable because more energetic CMEs are ejected in higher numbers during the maximum phase [Gopalswamy 2010]. Considering MCs also helps understand the geoeffectiveness based on the orientation of the flux rope with respect to the ecliptic and its variation with the solar cycle [Mulligan et al. 1998; 2000; Li and Luhmann 2004; Echer et al. 2005; Gopalswamy 2008; Kilpua et al. 2012; Szajko et al 2012; Lepping et al. 2015].

This study is concerned with all the MCs identified in in-situ data obtained at Sun-Earth L1 from May 1996 to December 2014. This period corresponds to the whole of cycle 23 and the beginning of the decay phase of cycle 24. Therefore, most of the inter-cycle variations considered in this work correspond to the first 73 months of cycles 23 and 24. Since the first 73 months include the rise and maximum phases of the two cycles, we also consider intra-cycle variations comparing the rise and maximum phases in the two cycles. We also investigate intra-cycle variations at a slightly finer scale by considering the annual averages of the MC parameters, somewhat similar to Dasso et al. [2012] and Lepping et al. [2015], although these authors consider different sets of parameters with some overlap with our parameters. In addition to the MC parameters, we also consider the properties of the ambient medium into which the MCs propagate and drive shocks. Our ultimate aim is to understand the mild space weather in cycle 24 as indicated by the low values of *Dst* index [Gopalswamy, 2012; Richardson 2013; Gopalswamy et al. 2014a]. Therefore, we consider the *Dst* index associated with both MC intervals and the compressed sheaths ahead of MCs. We use the *Dst* index as the primary indicator of geoeffectiveness, which is the ability of an interplanetary structure in causing a geomagnetic storm. According to Loewe and Prölss [1997], the storm level is indicated by Dst <-30 nT and an interplanetary structure resulting in such a storm level is considered to be geoeffective.

## 2. Observations

We started with the reported MC and MC-like (MCL) events during the first 73 months of each cycle (May 1996 to May 2002 in cycle 23; December 2008 to December 2014 in cycle 24). An MCL structure possesses most of the MC properties according to Burlaga et al. (1981) definition, except for the flux rope structure [Lepping et al. 2005]. For example of an ICME with low temperature and enhanced magnetic field but rotation in both $B_z$ and $B_y$ is considered an MCL. We eliminated such structures from consideration. The MCs of cycle 23 published before [Gopalswamy et al. 2008; Lepping et al. 2015] are used. We also examined some additional ICMEs listed in (http://wind.nasa.gov/index_WI_ICME_list.htm and http://www.srl.caltech.edu/ACE/ASC/DATA/level3/icmetable2.htm) to make sure that no MC was missed. For cycle 24, we examined the interplanetary CMEs (ICMEs) in online data identified MCs according to Burlaga et al.'s [1981] criteria.

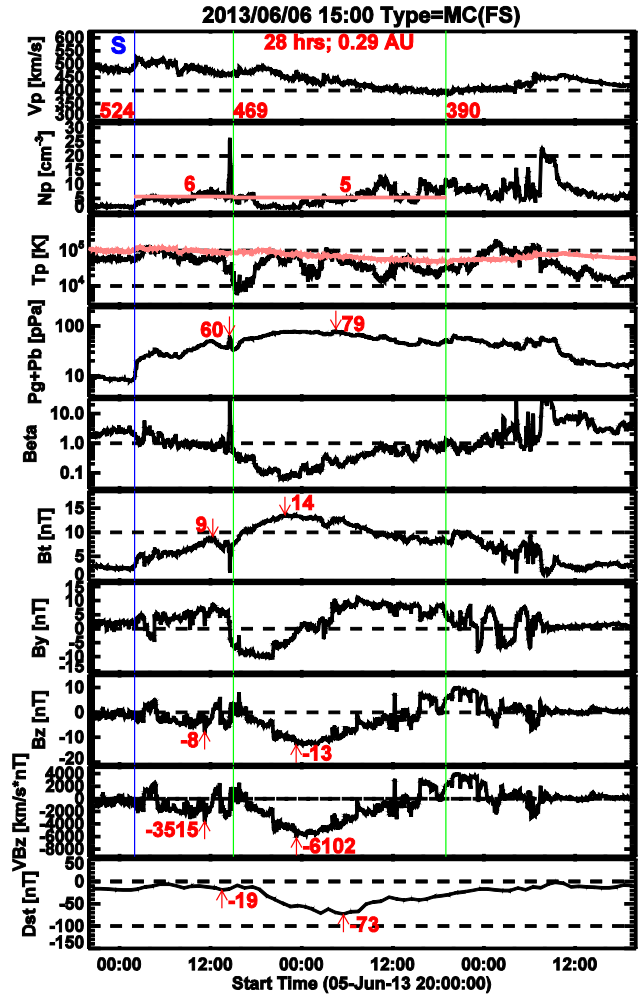

**Figure 1.** A set of plots used for identifying MCs. The MC in this plot was observed on 2013 June 6 at 15 UT. The MC lasts for 28 hours, ending at 19 UT on June 7 (marked by the two vertical green lines). The MC was driving a shock (S), denoted by the vertical blue line at 2 UT on June 6. The parameters plotted are (from top to bottom): plasma flow speed ($V_p$, km s$^{-1}$), proton density ($N_p$, cm$^{-3}$), proton temperature ($T_p$, K), total pressure ($P_t$, pPa) – the sum of magnetic ($P_b$) and gas ($P_g$) pressures, plasma beta, total magnetic field strength ($B_t$, nT), the Y-component of the magnetic field ($B_y$, nT), the Z-component of the magnetic field ($B_z$, nT), the product $VB_z$ (km s$^{-1}$·nT), and the $Dst$ index (nT). All the quantities but the $Dst$ index are from OMNI (omniweb.gsfc.nasa.gov) 1-min time resolution data. The $Dst$ index has a time resolution of 1 hour, as made available on line at the World Data Center, Kyoto (http://wdc.kugi.kyoto-u.ac.jp/dstdir/index.html). We used the final $Dst$ index if available; otherwise, we used the provisional or real-time data. The values indicated in red are the parameters that we compiled and used in the analyses. The data are also included as an electronic supplement (Table S1).

We also adjusted MC boundaries to be consistent with the Burlaga et al. [1981] definition of MCs and obtained several key parameters from the observations. Figure 1 shows an example of the plots made for each event and the parameters compiled from the plots. The plots were made using the OMNI data available online at NASA Goddard Space Flight Center (http://omniweb.gsfc.nasa.gov).We required that the proton temperature (Tp) falls below the expected solar wind proton temperature [Lopez and Freeman 1986] and/or the plasma beta is <1. We combined the Tp and beta signatures with the $B_y$ and $B_z$ components to identify the magnetic cloud type. In particular, we looked for events in which either $B_y$ or $B_z$ showed rotation, while the other stayed with the same sign throughout the MC interval. In Figure 1, for example, $B_z$ is negative (southward) throughout, while $B_y$ rotates from west to east (negative to positive in the Y-direction). This MC is therefore designated as fully south (FS). A variant with $B_y$ rotating from east to west with a similar $B_z$ profile will also be designated as an FS MC. Thus, the MC type is defined according to the axial direction and the direction of smooth rotation: north-south (NS), south-north (SN), fully south (FS) and fully north (FN). NS and SN clouds are known as bipolar clouds while FS and FN are termed unipolar clouds. NS and SN are low-inclination clouds with their axis close the ecliptic plane; FS and FN are high-inclination clouds with their axis perpendicular to the ecliptic plane [Gonzalez et al. 1990; Li and Luhmann 2004; Mulligan et al. 1998; 2000; Echer et al. 2005; Gopalswamy et al. 2008; Gopalswamy 2009]. The arrival time and the MC type are noted at the top of the plot in Fig.1.

The values indicated in red are the parameters we compiled. We also manually examined the plots to make sure the automatically identified numbers are not due to noise or other spikes. We also consulted *ACE* and *Wind* data directly if there is a data gap in the OMNI plots. For four events (Bastille Day 2000, Halloween 2003) the OMNI data were incomplete, and we used the data kindly provided by R. Skoug. These events were reported in Smith et al. [2001] and Skoug et al. [2004]. In Fig. 1, the flow speed is the highest at the leading edge of the cloud (469 km s$^{-1}$) and lowest at the tail end of the cloud (390 km s$^{-1}$). The average between these two numbers gives the central speed of the cloud, while the difference gives the expansion speed ($V_{exp}$). The solar wind speed at the time of the shock is taken as the sheath speed ($V_{sh}$). The ratio of the leading-edge speed to the central cloud speed is known as the expansion factor [$F_{exp}$, see e.g., Lepping et al. 2002]. There are many different ways of defining $V_{exp}$ [Owens et al. 2005; Gulisano et al. 2010], but we use the simplest form here as the difference speeds between the leading and trailing boundaries of the magnetic cloud. We also note the proton densities ($N_p$) in the sheath (6 cm$^{-3}$) and cloud (5 cm$^{-3}$) portions, averaged over the respective intervals. The peak field strength $B_t$ in the sheath and cloud portions are also recorded (9 and 14 nT, respectively) along with the total pressure in the two portions (60 and 79 pPa, respectively). Since we are interested in the geoeffectiveness of MCs, we also note the minimum value of the $B_z$ component in the sheath (-8 nT) and cloud (-13 nT) portions. For each data point, we multiplied the flow speed by $B_z$ and plotted the quantity $VB_z$ in units of nT.km s$^{-1}$. For the example in Fig. 1, the sheath and cloud have a minimum $VB_z$ of -3515 nT. km s$^{-1}$ and -6102 nT. km s$^{-1}$, respectively. All these parameters are tabulated for further analysis (data table S1), especially comparing their distributions between cycles 23 and 24.

Figure 2 shows the annual number of magnetic clouds, which shows a double peak, one in the rise phase and another in the maximum phase in both cycles [see also Riley et al. 2006]. The sunspot number and the annual number of frontside halo CMEs are also shown for reference.

Most of the magnetic clouds are associated with frontside halo CMEs because the latter head generally towards Earth. In the rise phase there are more MCs than the number of halos in both cycles. It is thought that these MCs not associated with halos are due to CMEs deflected toward the ecliptic plane by the large coronal-hole magnetic field [Gopalswamy et al. 2008]. Some MCs may also be associated with faint/undetectable halos or non-halos. In the maximum phase, there are more halos than the number of MCs. Many of these halos might have become non-cloud ICMEs, so we have not counted them. Since the polar filed disappears during the solar maximum, CME deflections similar to the rise phase do not occur, so some halos do not appear as MCs [the spacecraft is likely to go through the edge of the ICMEs, see Gopalswamy 2006]. In cycle 23, the ratio of number of halos to the number of MCs is very large, probably because there is an over count of halos due to misidentification of backside halos as frontside ones [Gopalswamy et al. 2015a] in addition to the lack of equatorward deflection noted above. It is known that there is a larger fraction of non-cloud ICMEs in the maximum phase [see e.g., Riley et al. 2006].

## 3. Analysis and Results

The primary purpose of this work is to compare the properties of MCs in cycles 23 and 24 in order to understand how the weak solar cycle 24 affected the geoeffectiveness of MCs. For this, we compare both sheath and cloud properties in the two cycles and their geoeffectiveness.

### 3.1 MC Occurrence Rate

Figure 2 compares the annual occurrence rate of MCs with that of frontside halo CMEs and the sunspot number (SSN). There is very little correspondence between SSN and the number of MCs. There is also a one-year lag between the minimum number of MCs (in 2008) and SSN minimum (in 2009). This is slightly different from Lepping et al. [2015], who reported that the two minima coincided. However, the minimum of MC number coincides with the minimum in the number of frontside halo CMEs. We see that the number of MCs did not decline significantly in cycle 24. In fact, there were 65 MCs in cycle 24 over the first 73 months (2008 December to 2014 December) with a very similar number (68) in cycle 23 over the corresponding epoch (1996 May to 2002 May). However, the average SSN (averaged over 73 months) was 75.98 in cycle 23 and 45.69 in cycle 24, indicating a decline of ~40% [Gopalswamy et al. 2015a]. In other words, the MC number did not decline as SSN did. Normalizing the MC rate to SSN, one can see that there were 1.4 MCs/SSN in cycle 24, compared to 0.88/SSN in cycle 23. The corresponding monthly rates were 0.019/SSN in cycle 24 compared to 0.012/SSN in cycle 23. Thus the MC abundance relative to SSN was higher in cycle 24, similar to the relative abundance of halo CMEs per month in this cycle (0.080/SSN versus 0.048/SSN in cycle 23); this is also evident from the relative abundance of frontside halos: 0.037/SSN (cycle 24) and 0.028/SSN (cycle 23) [Gopalswamy et al. 2015a]. MCs are expected to be a subset of frontside halos, but there are MCs that are associated with non-halo CMEs, which explains the difference between the rates of halo CMEs and MCs in Fig. 2. Moreover, MCs are generally associated with CMEs originating close to the disk center [Gopalswamy 2006], while halos can also originate at much larger central meridian distances [up to 90º - see e.g., Gopalswamy et al. 2007; 2010a; Cid et al. 2012; Gopalswamy et al. 2015a]. Furthermore, MCs and halo CMEs can also arise from non-spot regions, e.g., from quiescent filament regions [Gopalswamy et al. 2015b]. Finally, we note that the CME rate in the coronagraph field of view also did not decline significantly in cycle 24 as SSN did. This is not fully understood, although the weaker polar field [Petrie, 2013] and the anomalous expansion of CMEs [Gopalswamy et al. 2014a] have been proposed as possible explanations [Gopalswamy et al. 2015c].

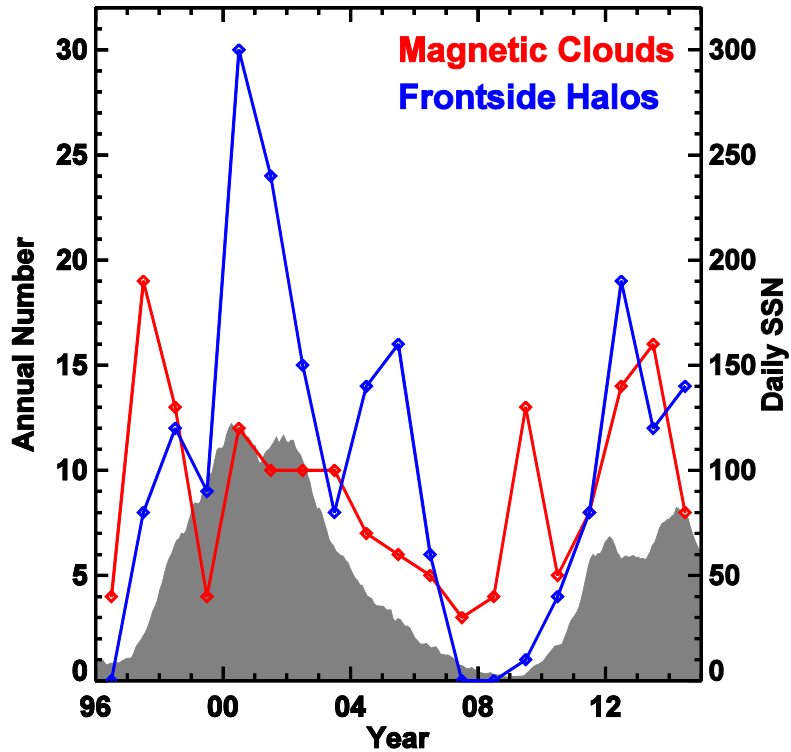

**Figure 2**. The annual number of magnetic cloud events as a function of time compared with front-side halo CMEs and the sunspot number (SSN).

### 3.2 Comparison of MCs between Cycles 23 and 24

In the following, we compare the MCs between cycles 23 and 24 to better understand their behavior in the two cycles. As noted in the beginning, the comparison will be made over the first 73 months of each cycle, unless stated otherwise. We also make intra-cycle comparison between solar-cycle phases. The rise and maximum phases in the two cycles are not of identical length. For example, cycle 23 started in May 1996 and the maximum phase ended in May 2002 (indicated by the completion of polarity reversal at the solar poles). The maximum phase started around the beginning of 1999 indicated by the arrival of polar crown filaments to a latitude of $60^o$ [Gopalswamy et al. 2003]. Based on similar considerations, we estimate the rise phase of cycle 24 to be in the interval December 2008 to July 2010 and the maximum phase as August 2010 to April 2014. While the first 73 months of cycle 23 corresponds to the rise + maximum phase in that cycle; it is not so in cycle 24 because the first 73 months of cycle 24 has a few months of decay phase (May 2014 to December 2014).

In order to assess the statistical significance of the differences in the physical parameters of MCs and the surrounding heliosphere in cycles 23 and 24, we used the Kolmogorov-Smirnov (KS) test the data sets that are compared. The comparison is made separately for MCs (Table 2) and sheaths (Table 3). The KS statistic $D$ is the maximum difference between the cumulative probabilities of two data sets. The critical value of $D$ depends on the number of data points $n$. For $n>50$, the critical value $D_c = 1.36n^{-1/2}$. For $n=65$, we have $D_c= 0.169$. For sheaths, the number of events is smaller ($n=40$), so $D_c =0.230$. Table 2 and 3 show the $D$ statistic obtained for the different MC and sheath parameters in addition to the $Dst$ index. In the last column of these tables, we have listed the probability ($P$) that the difference between the two data sets is by chance. If $P$ is <0.05, we reject the null hypothesis that the two distributions are the same.

Table 2. Kolmogorov Smirnov (KS) Test Results for MC Properties in Solar Cycles 23 and 24

| Param[a] | Cycle 23 ($n=68$) | | | Cycle 24 ($n=65$) | | | $R$[d] | $D$[e] | $P$[f] |
|---|---|---|---|---|---|---|---|---|---|
| | Mean | Conf. Int.[c] | Med | Mean | Conf. Int.[c] | Med | | | |
| $B_t$ | 16.54 | 14.87 to 18.20 | 14.55 | 12.33 | 11.23 to 13.43 | 12.20 | 0.75 | 0.3995 | 0.000 |
| $B_z$ | -10.90 | -12.43 to -9.37 | -10.35 | -7.80 | -8.93 to -6.67 | -7.50 | 0.72 | 0.2710 | 0.012 |
| $|B_z|$ | 13.33 | 11.93 to 14.72 | 11.75 | 10.23 | 9.24 to 11.21 | 9.70 | 0.77 | 0.2572 | 0.020 |
| $V$ | 473.9 | 439.9 to 507.8 | 445 | 402.1 | 384.5 to 419.7 | 399 | 0.85 | 0.2833 | 0.007 |
| $VB_z$ | -5119 | -6098 to -4139 | -4362 | -3078 | -3558 to -2599 | -2904 | 0.60 | 0.3373 | 0.001 |
| $\int VB_z$ | -2910 | -3429 to -2392 | -2450 | -1853 | -2172 to -1535 | -1595 | 0.64 | 0.2969 | 0.013 |
| $V^b_{exp}$ | 51.0 | 33.78 to 68.22 | 42.50 | 25.28 | 14.62 to 35.93 | 25.00 | 0.50 | 0.2425 | 0.033 |
| $F_{exp}$ | 1.053 | 1.038 to 1.068 | 1.050 | 1.032 | 1.020 to 1.045 | 1.030 | **0.98** | **0.1769** | **0.225** |
| $\zeta$ | 0.621 | 0.519 to 0.722 | 0.625 | 0.640 | 0.545 to 0.736 | 0.610 | **1.03** | **0.1157** | **0.867** |
| $P_t$ | 155.8 | 123.9 to 187.7 | 121.8 | 92.17 | 77.58 to 106.8 | 79.70 | 0.59 | 0.4283 | 0.000 |
| $N_p$ | 7.549 | 6.540 to 8.557 | 7.100 | 6.897 | 5.970 to 7.824 | 5.800 | **0.91** | **0.1441** | **0.464** |
| Size | 0.224 | 0.202 to 0.246 | 0.230 | 0.174 | 0.151 to 0.197 | 0.170 | 0.78 | 0.2704 | 0.012 |
| $\Delta t$ | 21.99 | 19.56 to 24.42 | 21.45 | 19.19 | 16.68 to 21.70 | 18.00 | **0.87** | **0.1762** | **0.229** |
| $D_{st}$ | -65.54 | -78.29 to -52.79 | -56.00 | -33.37 | -42.27 to -24.47 | -22.00 | 0.51 | 0.3318 | 0.001 |

[a]Units of the parameters: $B_t$, $B_z$ in nT; MC speed $V$, and expansion speed $V_{exp}$ in km s$^{-1}$; $VB_z$ in nT. km s$^{-1}$; $\int VB_z$ is normalized over the interval of southward $B_z$ and the units are nT. km s$^{-1}$; $F_{exp}$, $\zeta$-dimensionless; $P_t$ in pPa; $N_p$ in cm$^{-3}$; size in AU; $\Delta t$ in hours; $Dst$ in nT. [b]Only positive expansion speeds are used: 55 in cycle 23 and 48 in cycle 24; [c]the 95% confidence interval of the actual mean; [d]the ratio of means (cycle 24 to cycle 23); [e]the $D$ statistic is the maximum difference in the cumulative probabilities of the quantities in cycles 23 and 24. [f]The probability that the high $D$ value is due to chance.

Table 3. Kolmogorov Smirnov (KS) Test Results for Sheath Properties in Solar Cycles 23 and 24

| Param[a] | Cycle 23 ($n=55$) | | | Cycle 24 ($n=40$) | | | $R$[c] | $D$[d] | $P$[e] |
|---|---|---|---|---|---|---|---|---|---|
| | Mean | Conf. Int | Median | Mean | Conf. Int. | Median | | | |
| $B_t$ | 19.18 | 15.76 to 22.59 | 15.00 | 13.24 | 11.19 to 15.28 | 11.55 | 0.69 | 0.3318 | 0.009 |
| $B_z$ | -14.23 | -17.26 to -11.19 | -11.90 | -8.92 | -11.17 to -6.679 | -7.7 | 0.63 | 0.3727 | 0.002 |
| $|B_z|$ | 15.87 | 12.82 to 18.92 | 13.10 | 11.47 | 9.584 to 13.37 | 9.7 | 0.72 | 0.3250 | 0.011 |
| $V$ | 539.5 | 494.4 to 584.7 | 484 | 447.4 | 417.1 to 477.8 | 426.5 | 0.83 | 0.2932 | 0.029 |
| $VB_z$ | -8101 | -10840 to -5363 | -5600 | -3988 | -5251 to -2726 | -3040 | 0.49 | 0.3659 | 0.003 |
| $\int VB_z$ | -3260 | -4381 to -2131 | -2370 | -1634 | -2060 to -1208 | -1245 | 0.50 | 0.3364 | 0.008 |
| $P_t$ | 272.9 | 182.1 to 363.6 | 168.8 | 135.2 | 96.18 to 174.3 | 93.60 | 0.50 | 0.3432 | 0.006 |
| $N_p$ | 15.25 | 12.85 to 17.65 | 13.20 | 12.90 | 10.85 to 14.95 | 11.35 | **0.85** | **0.1932** | **0.319** |
| $\Delta t$ | 11.19 | 9.40 to 12.98 | 10.70 | 9.635 | 7.948 to 11.32 | 9.2 | **0.86** | **0.1273** | **0.821** |
| $D_{st}$ | -55.18 | -71.45 to -38.92 | -37.00 | -22.75 | -33.63 to -11.87 | -12.00 | 0.41 | 0.4409 | 0.000 |

[a]Units of the parameters: $B_t$, $B_z$, $|B_z|$ in nT; sheath speed $V$ in km s$^{-1}$; $VB_z$ in nT. km s$^{-1}$; $\int VB_z$ is normalized over the interval of southward $B_z$ and the units are nT. km s$^{-1}$; $P_t$ in pPa; $N_p$ in cm$^{-3}$; size in AU; $\Delta t$ in hours; $Dst$ in nT; [b]the 95% confidence interval of the actual mean; [c]the ratio of means (cycle 24 to cycle 23); [d]the $D$ statistic is the maximum difference in the cumulative probabilities of the quantities in cycles 23 and 24. [e]The probability that the high $D$ value is due to chance.

[a]Units of the parameters: $B_t$, $B_z$ in nT; MC speed $V$, and expansion speed $V_{exp}$ in km s$^{-1}$; $VB_z$ in nT. km s$^{-1}$; $\int VB_z$ is normalized over the interval of southward $B_z$ and the units are nT. km s$^{-1}$; $F_{exp}$, $\zeta$ -dimensionless; $P_t$ in pPa; $N_p$ in cm$^{-3}$; size in AU; $\Delta t$ in hours; $Dst$ in nT. [b]Only positive expansion speeds are used: 55 in cycle 23 and 48 in cycle 24; [c]the 95% confidence interval of the actual mean; [d]the ratio of means (cycle 24 to cycle 23); [e]the $D$ statistic is the maximum difference in the cumulative probabilities of the quantities in cycles 23 and 24. [f]The probability that the high $D$ value is due to chance.

Tables 2 and 3 show the mean, median, and the 95% confidence interval of the mean for each quantity compared between the two cycles. The ratio of means ($R$) shows the extent to which the parameters changed in cycle 24. The quantity (1-$R$) expressed in percentages is the reduction of the quantity in cycle 24 relative to cycle 23. This reduction percentage is used throughout the paper. Another way to see whether the difference of a quantity between cycles 23 and 24 is significant or not is to compare the 95% confidence intervals of the mean values. For example, the mean MC durations ($\Delta t$) in Table 2 are 21.99 h (cycle 23) and 19.19 (cycle 24). The 95% confidence intervals are 19.56 to 24.42 (cycle 23) and 16.68 to 21.70 (cycle 24), which heavily overlap and hence the duration in the two cycles are not very different. This is indicated by the large ratio $R=0.87$ and the small $D$ value, 0.1762. The fact that $P$ value 0.229 is >> 0.05 indicates that the null hypothesis cannot be rejected (i.e., the difference between the two distributions is not significant). For parameters such as $\Delta t$, we have indicated $R$, $D$, and $P$ values in bold face.

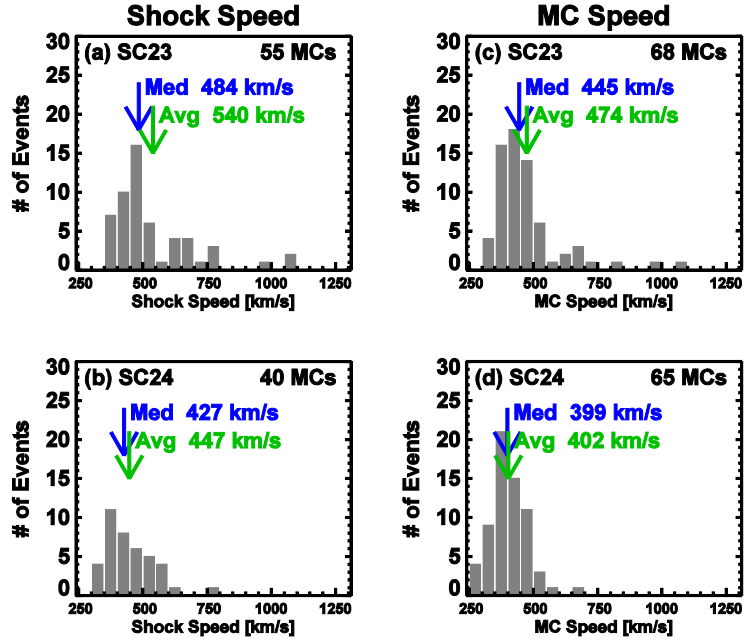

**Figure 3**. Comparison between shock (a,b) and MC (c,d) speeds for solar cycle 23 (top) and 24 (bottom). The mean (Avg) and median (Med) speeds are noted on the plots. Only 55 MCs in cycle 23 (81%) and 40 MCs in cycle 24 (62%) were driving shocks.

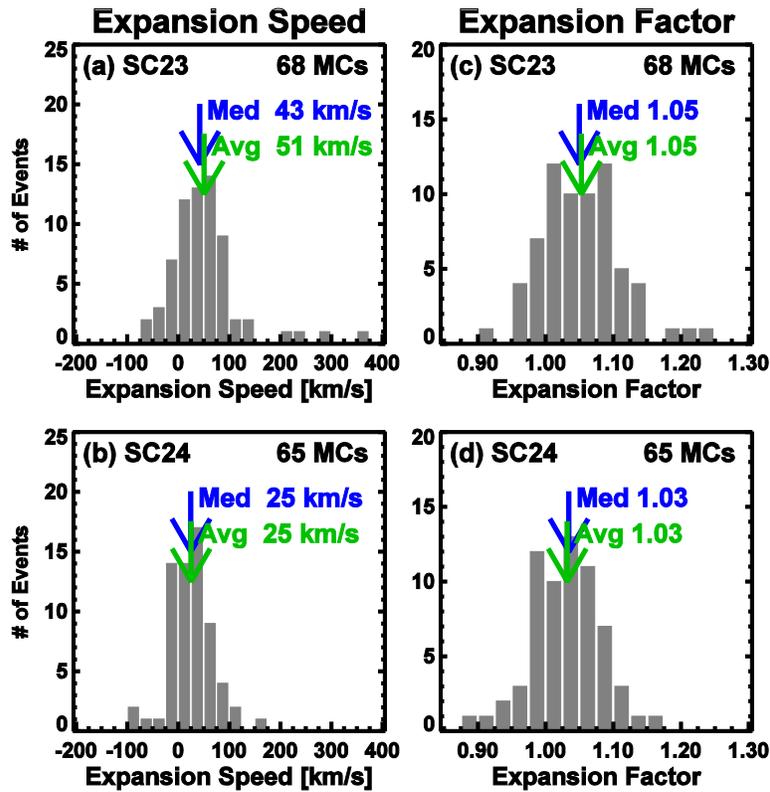

**Figure 4**. The expansion speed (a,b) and expansion factor (c,d) for solar cycles 23 (top) and 24 (bottom). The expansion speeds in the two cycles are significantly different, while the expansion factors remain the same.

### 3.2.1 Magnetic Cloud Types

It is well known that bipolar MCs show a 22-year variation in the predominance of one type over the other [Echer et al. 2005 and references therein]. The predominance begins right after the completion of polarity reversal at the close of the maximum phase of a cycle and ends at the subsequent polar reversal. This corresponds to half of the 22-year cycle. From the declining phase of cycle 22 to the maximum phase of cycle 23, SN MCs were predominant. After the polarity reversal in May 2002 in cycle 23, Gopalswamy et al. [2008] showed that the number of NS type clouds started increasing [their figure 4; see also figure 3 in Li et al. 2011]. The predominance of NS MCs continued until April 2014, when the polarity reversal occurred in cycle 24. There were a total of 108 MCs from June 2002 to April 2014 and 44 (or 42%) were NS MCs (listed in the supplementary data table S1). Starting in May 2014, SN clouds started occurring more frequently again, confirming the 22-year cycle: out of the five MCs observed between May and December 2014, three were bipolar and of SN type; the remaining two were unipolar (one was FN and the other FS). This is also consistent with the 22-year sign change of the solar polar field. In essence, the leading polarity of the bipolar MC is the same as that of the global (bipolar) solar field [Echer et al. 2005].

Table 1. MC types in the rise and maximum phases of cycles 23 and 24

| MC Type | Cycle 23 | | | Cycle 24 | | |
|---|---|---|---|---|---|---|
| | RISE (5/96 - 12/98) | MAX (1/99 – 5/02) | Total (5/96 - 5/02) | RISE (12/08- 7/10) | MAX (08/10- 4/14) | Total (12/08- 4/14) |
| SN | **16** | **15** | **31 (46%)** | 2 | 10 | 12 (20%) |
| NS | 4 | 7 | 11 (16%) | **12** | **13** | **25 (40%)** |
| FS | 11 | 4 | 15 (22%) | 2 | 10 | 12 (20%) |
| FN | 5 | 6 | 11(16%) | 4 | 8 | 12 (20%) |
| Total | 36 | 32 | 68 (100%) | 27 | 33 | 61 (100%) |

Table 1 shows the number of MCs of different types in the rise and maximum phases of cycles 23 and 24. The largest subsets are SN MCs in cycle 23 and NS MCs in cycle 24 with similar fractions of all MCs (46% and 40%, respectively – shown in bold in Table 1). Counting only the bipolar MCs (SN+NS), we see that SN MCs dominate in cycle 23 (31 SN MCs out of the 42 bipolar MCs or 74%); similarly, NS MCs dominate in cycle 24 (25 NS MCs out of the 37 bipolar or 68%). It must be noted that the combined rise and maximum phases correspond to only about half of the period during which one type of bipolar MCs dominates. Table 1 also shows that SN MCs occur in equal abundance in the rise (16) and maximum (15) phases. A similar trend can be seen for the NS MCs in cycle 24 (rise: 12 and maximum: 13). The FN MCs do not show any solar cycle trend, while the FS MCs seem to be dominant in the rise phase of cycle 23 and maximum phase of cycle 24. Further investigation is needed for a better understanding of the unipolar MCs because their numbers are too small to perform statistical analyses. Comparing the MC types in cycles 23 and 24, we conclude that the predominant MC type follows the expected cyclical behavior (predominance of SN and NS MCs during odd and even cycles, respectively) despite the weakness of cycle 24 in terms of SSN.

### 3.2.2 Plasma and Magnetic Field Properties

The state of the heliosphere in cycle 24 is known to be significantly weak: the proton density, proton temperature, magnetic field strength, total pressure, and Alfven speed all diminished with respect to those in cycle 23 [McComas et al. 2013; Gopalswamy et al. 2014a]. It has also been found that white-light CMEs expand anomalously in cycle 24, resulting in a larger width within the coronagraph field of view (FOV) than in cycle 23. Due to expansion, narrow CMEs tend to appear like normal CMEs, thereby reducing the average CME mass in cycle 24 ($1.1 \times 10^{15}$ g) compared to cycle 23 ($3.2 \times 10^{15}$ g) [Gopalswamy et al. 2015c]. It is of interest to know how these changes reflect in the properties of MCs because MCs are evolved forms of CMEs observed in white light. In particular, the magnetic field in CMEs near the Sun is currently not measurable, although the strength and orientation of the magnetic field are important factors in causing geomagnetic storms. Magnetic properties of MCs are well measured at 1 AU, so we make use of the in-situ observations in comparing MC properties between the two cycles to understand the poor geoeffectiveness of CMEs in cycle 24. In comparing various distributions of physical parameters in the two cycles, we use the mean values of the distributions throughout the paper.

**Sheath and MC speeds:** Figure 3 compares the distributions of shock and MC speeds for the two cycles. The shock speed is taken as the sheath speed. In both cycles, the shock speed is greater than the leading-edge speed of MCs by about 10%. Both shock and MC speeds in cycle 24 are smaller than the corresponding values in cycle 23. The most probable value is at a lower bin in cycle 24 for the shock and MC speeds. The average MC speed in cycle 24 is about 15% lower than that in cycle 23, with a similar reduction in the shock speeds (~17%). Tables 2 and 3 show that the speed difference is statistically significant. This result is different from what was observed at the Sun: the white-light CME speeds were roughly the same in the two cycles, but the cycle-24 CME widths were larger on the average. A possible explanation is that the wider CMEs in a slower background solar wind [McComas et al. 2013] are subject to a larger drag force in the interplanetary medium, resulting in a lower MC speed at 1 AU. The number of shock-driving MCs is also smaller in cycle 24 (62% vs. 81% in cycle 23).

**Expansion Speeds and Expansion Factors:** We also examined the expansion properties of MCs based on their leading ($V_L$) and trailing ($V_T$) edge speeds. The expansion speed is taken as $V_{exp} = V_L - V_T$. When $V_L > V_T$ the MC is expanding. Most of the MCs show expansion at 1 AU (56 out of 68 or 82% in cycle 23 compared to 48 out of 65 or 74% in cycle 24). In the non-expanding cases ($V_L \leq V_T$), a corotating interaction region usually followed, which means the MC flow is perturbed and hence does not follow the typical expansion pattern [see e.g., Gulisano et al. 2010; Dasso et al. 2012]. We also computed the expansion factor defined as $F_{exp} = V_L / V_C$, where the central speed of the MC $V_c = (V_L + V_T)/2$ [see Lepping et al. 2002 for details]. Figure 4 shows the distributions of expansion speeds and the expansion factors. The average $V_{exp}$ of ~25 km s$^{-1}$ in cycle 24 is ~50% smaller than the 51 km s$^{-1}$ in cycle 23 (see Table 2). The $V_{exp}$ distribution is also narrower in cycle 24: the range is restricted to <200 km s$^{-1}$ in cycle 24, while there are a few values up to 400 km s$^{-1}$ in cycle 23. The $V_{exp}$ was typically a small fraction of $V_L$ (10.8% in cycle 23 and 6.3% in cycle 24).

$F_{exp}$ ranged from 0.85 to 1.30 in cycle 24 with a similar range in cycle 23 (0.90 to 1.35). The average values of the expansion factor are also similar: 1.032 (cycle 24) and 1.053 (cycle 23) as can be seen in Fig. 4. These values are similar to those reported by Lepping et al. [2002] for 27 MCs (range: 0.92 to 1.26). $F_{exp}$ is close to unity because of the small expansion speed compared to the leading-edge speed of MCs. The difference between the two $F_{exp}$ distributions in Fig. 4 is not statistically significant (see Table 2). $F_{exp}$ is similar in cycles 23 and 24 because both the leading-edge speed and the MC central speed declined in cycle 24, keeping their ratio roughly the same.

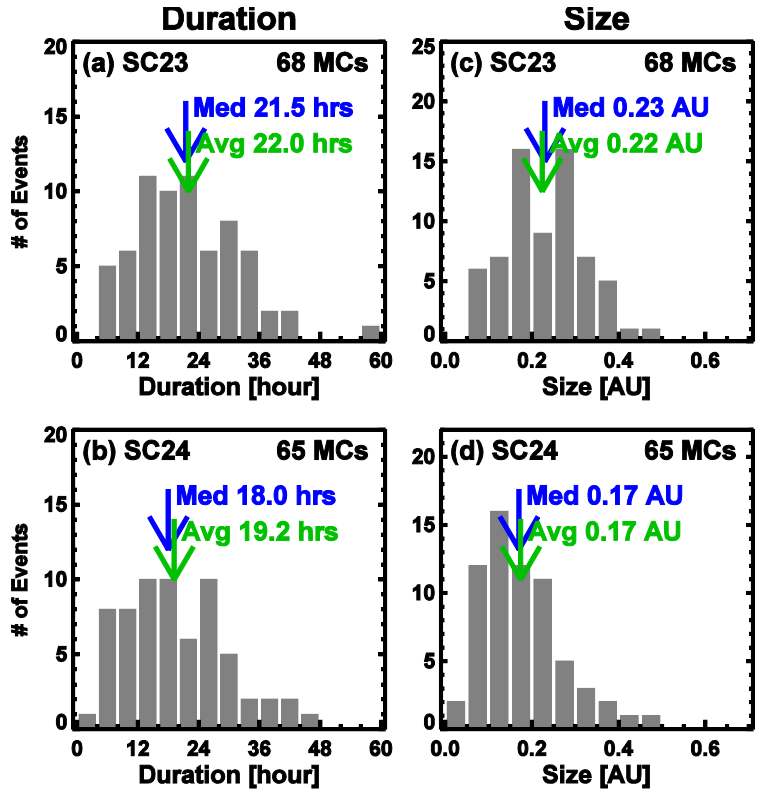

**Figure 5**. The duration (left) and size (right) of the magnetic clouds at 1 AU

**MC Duration and Size:** The magnetic cloud duration ($\Delta t$) is simply the time difference between the front and back boundaries of the cloud. The cloud size was determined by integrating the speed with time between the two MC boundary times. Figure 5 shows the distributions of $\Delta t$ and size of the MCs for the two cycles. The MC durations are very similar between cycles 23 and 24. The 95% confidence intervals of the means in cycles 23 and 24 had significant overlap, suggesting that the distributions are not different (see Table 2). On the other hand, the MC sizes were significantly smaller in cycle 24, although the ranges were similar. The average size dropped by 23% in cycle 24. One possibility is that the anomalous expansion of CMEs near the Sun makes narrow CMEs appear like normal CMEs in white light. This means a significant fraction of cycle-24 flux ropes would have started out as smaller ones. This is also inferred from the lower CME mass in cycle 24 on average.

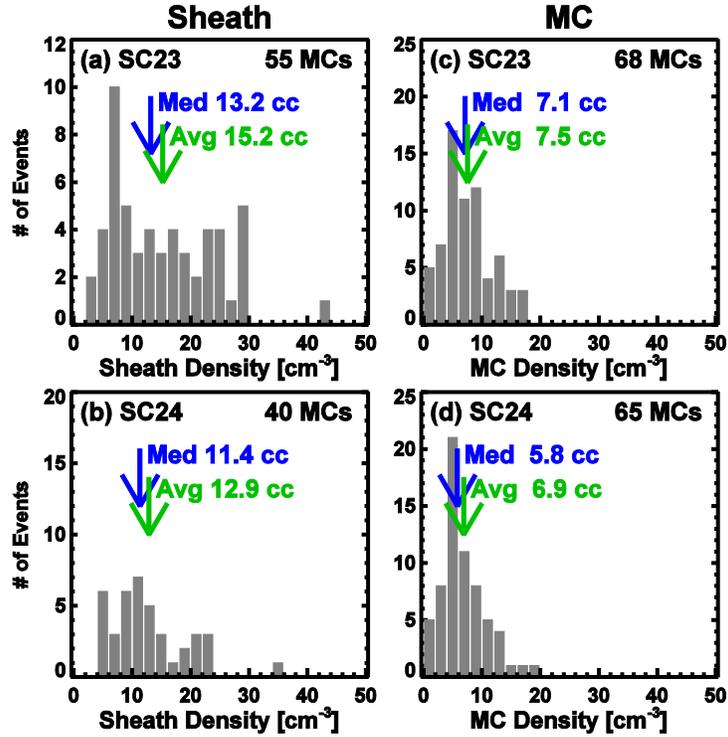

**Figure 6.** Comparison between sheath (a,b) and MC (c,d) densities for solar cycle 23 (top) and 24 (bottom). The mean (Avg) and median (Med) speeds are noted on the plots.

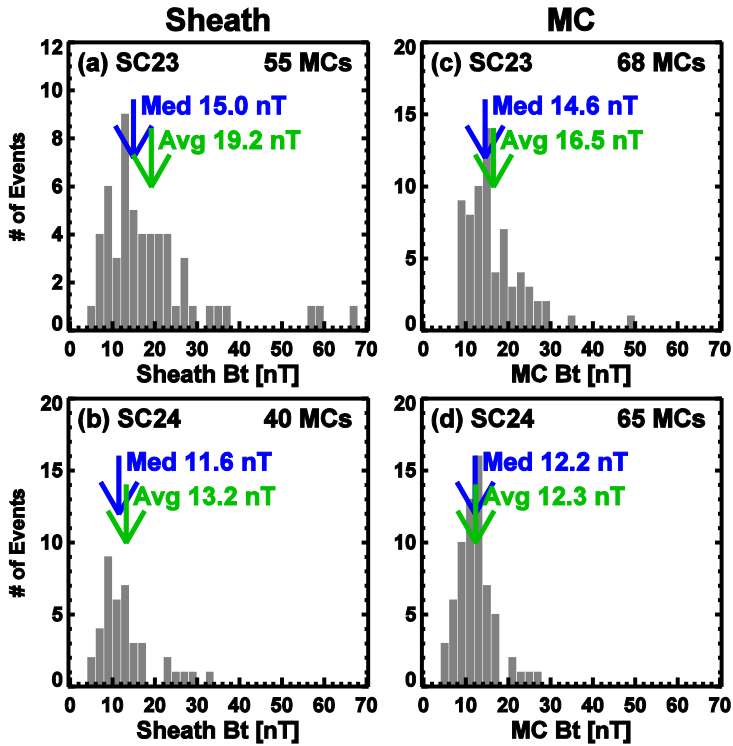

**Figure 7**. Comparison between sheath (a,b) and MC (c,d) peak field strengths ($B_t$) for solar cycle 23 (top) and 24 (bottom). The mean and median $B_t$ values are noted on the plots.

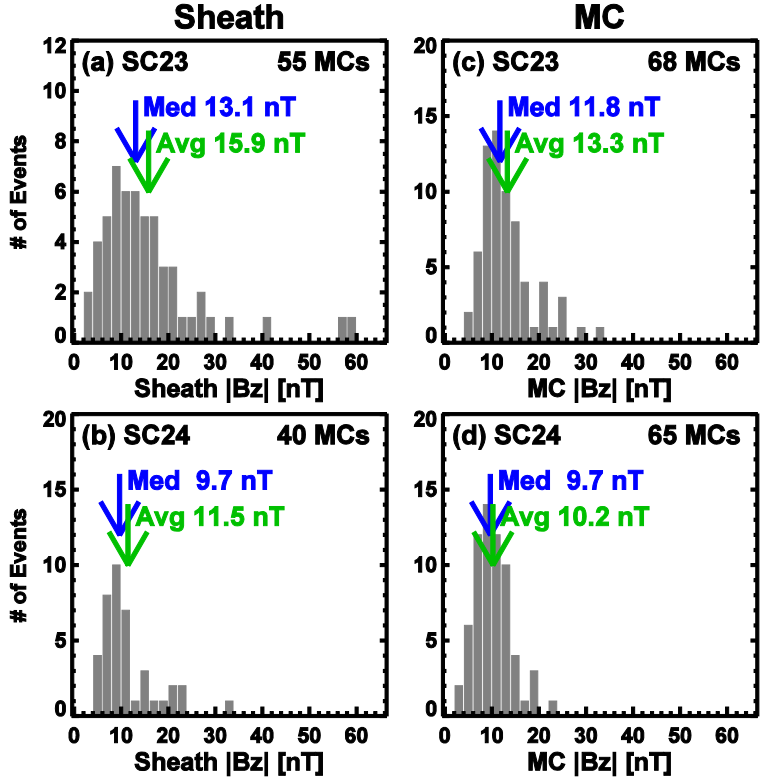

**Figure 8**. The absolute maximum values of $B_z$ in the sheath (a,b) and MC (c,d) portions.

**Proton Densities:** The sheath density is essentially the heliospheric increased density in the shock downstream, ahead of the MC. The density inside the magnetic cloud is determined by the initial density at the Sun and the way the MC expands. Leitner et al. [2007] showed that the density inside MCs falls off more rapidly with distance than the ambient solar wind because of the MC expansion. Figure 6 compares the density distributions in the sheath and cloud portions for the two cycles. The sheath and MC densities are not too different in the two cycles. The sheath and cloud mean densities fall by 15% and 9%, respectively. However, KS test showed that the difference in the distributions between the two cycles is not significant (see Table 2). In fact, the 95% confidence intervals for the means overlap heavily suggesting that the difference may be by chance.

**Magnetic Field Strengths:** Figure 7 compares the peak magnetic field strength $B_t$ in the sheath and cloud portions for cycle 23 and 24. Once again the most probable $B_t$ values in cycle 24 are at a lower bin for cycle 24 MCs and sheaths. The distributions are significantly different with the cycle-24 mean $B_t$ dropping by 31% and 25% in the sheath and MC portions. The KS test results in Tables 2 and 3 show that that these drops are significant with the $D$ statistic much greater than the critical values (0.3993 for MCs and 0.3318 for sheaths). The probability $P$ that these high values of $D$ are due to chance is close to 0.0 (see Table 2).

It must be noted that $B_t$ in the MC and sheath have different sources. In MCs, $B_t$ is related to the source region magnetic field, while the one in the sheath is the compressed heliospheric magnetic field ahead of MCs. Both show a decline with respect to the cycle 23 values. The heliospheric field is diminished in cycle 24, so MHD compression in the sheath can increase $B_t$ only by a factor <4. On the other hand, the decrease of $B_t$ in MCs is likely due to the increased expansion of cycle 24 CMEs near the Sun, which weakens the magnetic field strength.

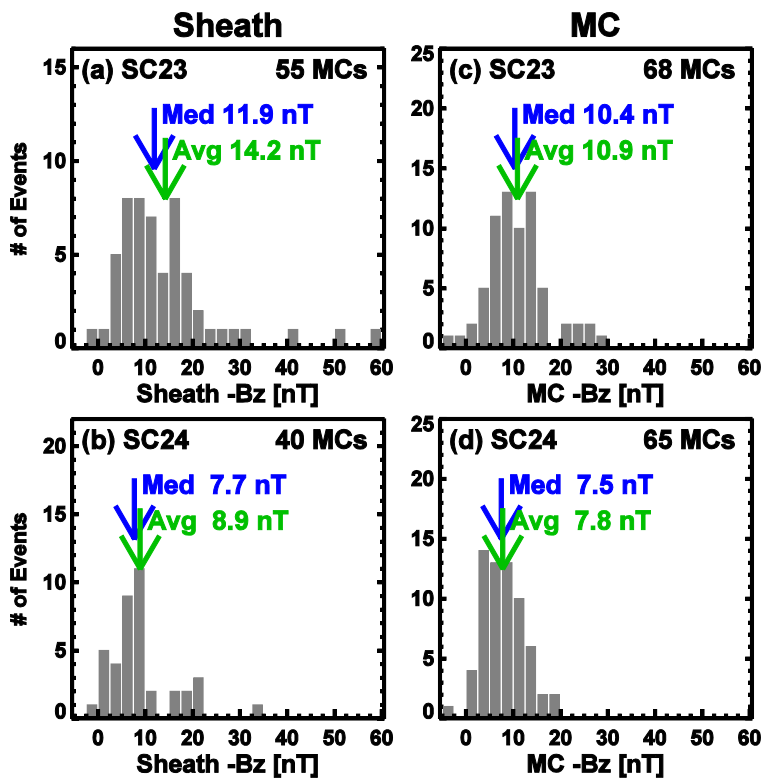

**Figure 9**. The minimum values of $B_z$ in the sheath (a,b) and MC (c,d) portions. The small number of positive $B_z$ are due to the FN type clouds. A few sheaths also did not have negative $B_z$.

Figure 8 shows the maximum of the absolute values of $B_z$ in the sheath and cloud intervals. Since $B_z$ is a part of $B_t$, it should reflect the behavior of $B_t$. In the MC part, $|B_z|$ is lower in cycle 24 by 23%, similar to the 25% in $B_t$. The reduction in the sheath part is also similar: 28% for $|B_z|$ and 31% for $B_t$. Thus the reduction in the magnetic content of MCs is larger than the reduction in density between the two cycles noted in Fig. 6.

The geoeffectiveness of CMEs depends not only on the magnitude of $B_z$, but also the sign. $B_z$ needs to be negative in either or both of the cloud and sheath portions for a geomagnetic storm to occur. Therefore, we identified the minimum value of $B_z$ in the sheath and cloud portions. Figure 9 shows the distributions of minimum $B_z$ values for the two cycles. The mean values of the distributions (sheath and cloud portions) are significantly smaller in cycle 24. The reduced $B_z$ combined with the slower MCs in cycle 24 are likely to reduce their geoeffectiveness as we shall see next.

We computed the speed ($V$) – $B_z$ product at each time step in the data and identified the minimum value (the largest negative value in most cases). The resulting $VB_z$ values are plotted in Fig. 10 for the two cycles. Even though the values corresponding to the FN MCs are not significant because they represent noise, we have shown them here for completeness and the average and median values of the distributions do not change if these values are removed. The plots show that the $VB_z$ distributions in cycle 24 drop off rapidly after $-1\times10^4$ nT. km s$^{-1}$. The largest negative value in cycle 24 is -7.68×10³ km s⁻¹.nT compared to -2.54×10⁴ km s⁻¹.nT. In the sheaths, the average $VB_z$ declines by 51% while the decline is ~40% in the clouds. Both of these drops are statistically significant as the KS test shows (see Tables 2 and 3). The cycle-24 MC values are normally distributed, while the cycle-23 values are not because of the long tail. The tail values beyond -1×10⁴ nT. km s⁻¹ in cycle 23 are responsible for geomagnetic storms stronger than -100 nT.

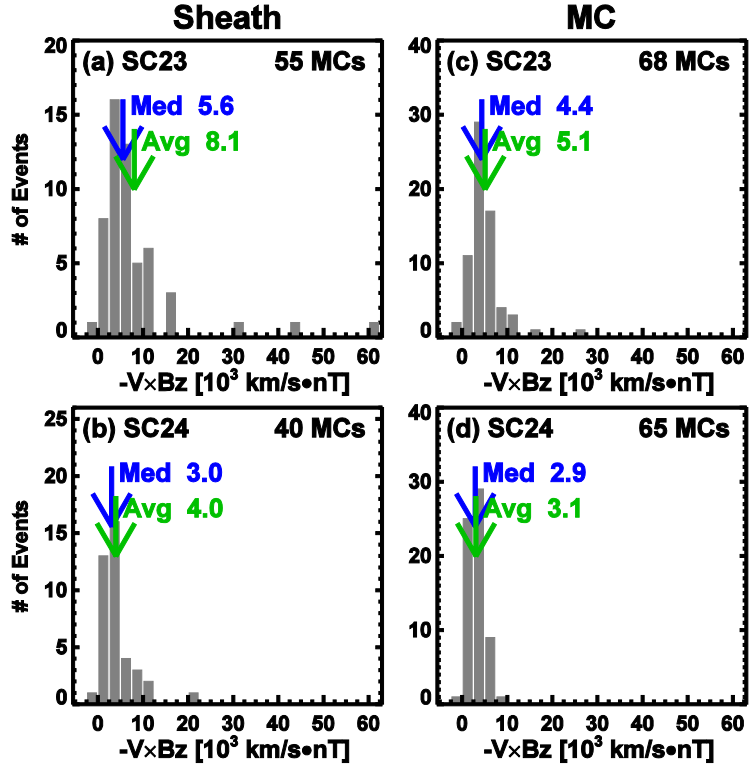

**Figure 10**. Distributions of $VB_z$ in the sheath (a,b) and cloud (c,d) portions for cycles 23 (top) and 24 (bottom). There are not many $VB_z$ values < -10⁴ km s⁻¹.nT in cycle 24. The positive values in a few cases are due to FN clouds and in sheaths that do not have negative $B_z$.

Since our main interest is to compare the minimum value of *Dst* following each MC (and their sheath), we considered only the minimum Details on the evolution of *Dst* can be followed better when one uses the time integral of the *y*-component of the electric field, viz., ∫$VB_z dt$ [see e.g., Kane 2010, Ontiveros and Gonzalez-Esparzza 2010; Yermolaev et al. 2010; Weigel 2010; Nikolaeva et al. 2013; 2015]. We computed ∫$VB_z dt$ and (1/*T*) ∫$VB_z dt$ where *T* is the interval of negative $B_z$. We found that the drop in (1/*T*) ∫$VB_z dt$ in cycle 24 is nearly the same as minimum $VB_z$ for MCs (36%) and sheaths (50%). Recall that the corresponding $VB_z$ values are 40% (MC) and 51% (sheath). KS test results given in Tables 2 and 3 show that the difference in (1/*T*) ∫$VB_z dt$ between cycles 23 and 24 are statistically significant.

### 3.2.3 The Speed- Magnetic Field Relationship

The importance of CMEs for geomagnetic storms stems from the fact that CME flux rope structure introduces an out-of-the-ecliptic component of the heliospheric magnetic field, which is otherwise in the ecliptic plane, except for Alfvenic fluctuations [e.g., Wilson 1987]. The flux-rope field derives from the source region on the Sun, whether an active region or a filament region. The three types of magnetic clouds, viz., SN, NS, and FS all have southward field somewhere within the duration of the MC. On the other hand, FN type MCs do not possess a southward field (the axis is north pointing), so they do not produce a significant geomagnetic response. The sheaths of all the four types of clouds are likely to contain southward $B_z$, and hence are likely to cause geomagnetic storms. The geoeffectiveness of MCs and sheaths will be discussed in the next session. Here we discuss the relation between speed and the magnetic field strength in the MC and sheath.

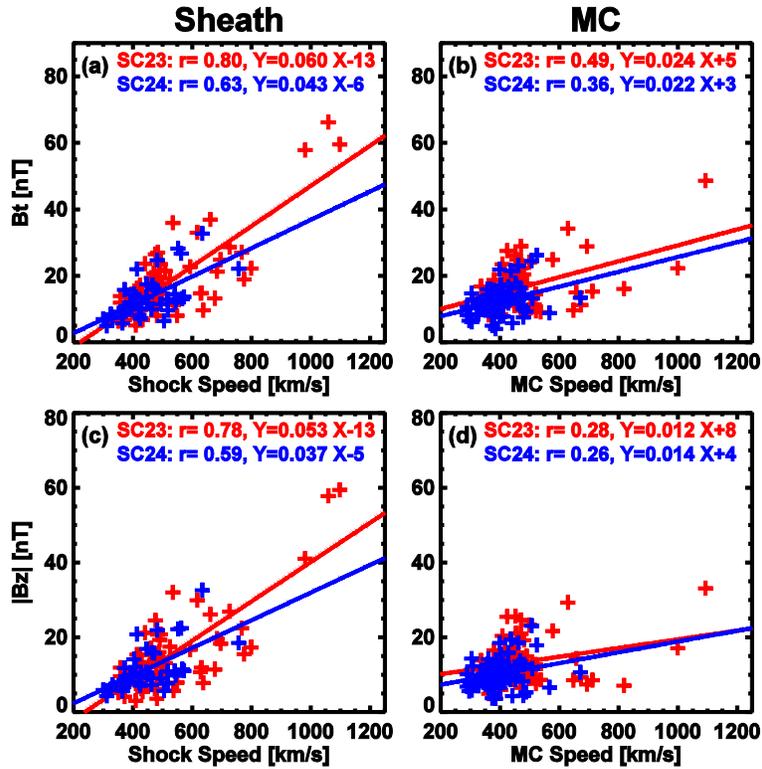

**Figure 11**. Scatter plot between magnetic field strength and speed in the sheath (left) and cloud (right) portions for the two cycles. The regression lines and the correlation coefficients (r) are also shown on the plots.

Gonzalez et al. [1998] were the first to point out that the peak field strength and speed in MCs are correlated (r=0.75) for a set of 30 MCs. A much smaller correlation (r= 0.35) was obtained for a larger set of 149 MCs [Echer et al. 2005]. For about 100 MCs during cycle 23, Gopalswamy et al. [2008] found this correlation to be valid (r=0.56) over a wider range of speeds. On the other hand, Owens et al. [2005] reported that the magnetic field - speed correlation is significant only in the sheath region ahead of magnetic clouds, and not within the clouds themselves. Gopalswamy et al. [2008] used the average MC speed in order to eliminate the expansion speed of the clouds, but the correlation was still present. Since the expansion speed is a small fraction of the MC speed (see Figs. 3 and 4), we use the leading-edge speed in performing correlation analysis. Since we are interested in comparing MCs during the first 73 months of cycles 23 and 24, we have a smaller number of MCs, but still more than twice the number used by Gonzalez et al. [1998] and Owens et al. [2005]. The scatter plots in Figure 11 show that the $V$-$B_t$ and $V$-$|B_z|$ correlations are high in the sheath for both cycles, confirming the results of Owens et al. [2005]. The $V$-$B_t$ correlations are also significant in the cloud portions: r=0.43 in cycle 23 and 0.36 in cycle 24. Similar correlations were also reported by Echer et al. [2005]. These correlations are statistically significant because the critical value of the Pearson correlation coefficient for $P$=0.05 is 0.255 for MCs and 0.282 for sheaths. The only case with no significant correlation is $V$-$|B_z|$ for MCs in cycle 23 (r=0.20) and 24 (r=0.26). On closer examination, we noticed that the $V$-$B_t$ and $V$-$|B_z|$ correlations have two branches. The higher values in both branches are from the maximum phase of cycle 23. It is not clear why these MCs from the maximum phase have low field strength. When we excluded these high-speed, low-field values, the correlation coefficients increased significantly: 0.66 and 0.39 for $V$-$B_t$ in cycles 23 and 24, respectively. Similarly, the $V$-$|B_z|$ correlations also improved to 0.39 and 0.26, respectively for cycle 23 and 24. It is not clear why the data points organize in two branches (with the lower branch coming from the maximum phase) and requires further investigation. It may be simply an issue of statistics because the number of points is small.

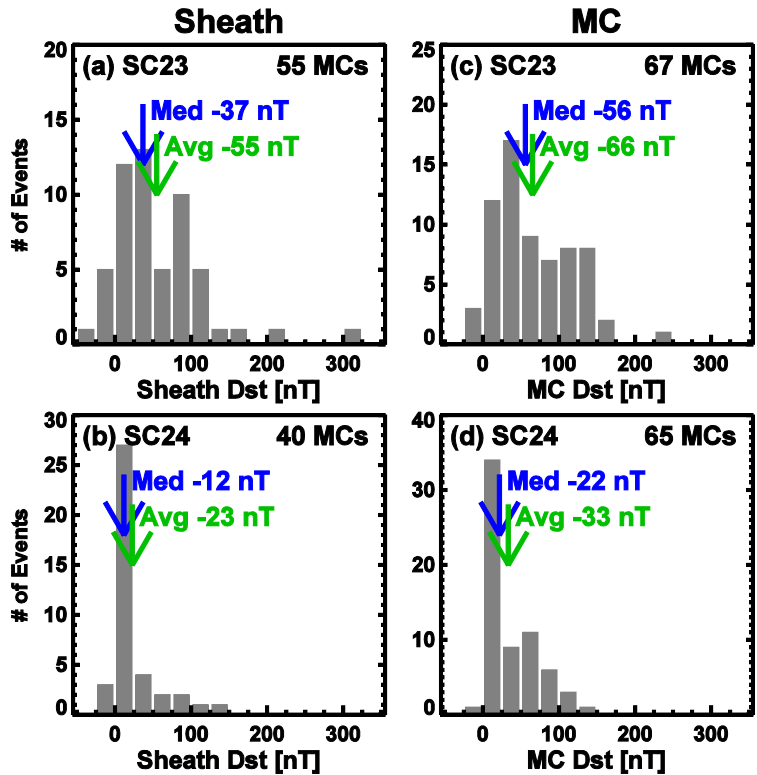

**Figure 12**. Comparison between sheath (a,b) and MC (c,d) *Dst* values for solar cycle 23 (top) and 24 (bottom). The mean and median *Dst* values are noted on the plots.

### 3.2.4 Geoeffectiveness

We identified the minimum *Dst* values that can be attributed to the sheath and cloud portions for each MC. For FN MCs, the *Dst* index for the cloud portion is zero, but there may be small values in some cases due to noise. In cycle 23, the Bastille Day 2000 super storm was mainly caused by the sheath, but there may be a contribution of unknown level from the MC because it was a SN cloud. For this reason, we do not use any *Dst* value for the MC part. This is the reason for the 67 *Dst* values for MCs, rather than 68. Figure 12 compares the *Dst* distributions in sheath and cloud portions for the two cycles. It is clear that the cycle-24 distributions are narrower with a limited range of *Dst* values. The average values of the *Dst* distributions are significantly smaller in cycles 24 for both sheaths (by 59%) and MCs (by 49%). KS test results presented in Tables 2 and 3 show that the difference in *Dst* between cycles 23 and 24 are statistically significant. Note that the drop in the average *Dst* values between the two cycles is similar to the drop in $VB_z$ values discussed above because the minimum value of *Dst* is correlated with the minimum value of $VB_z$.

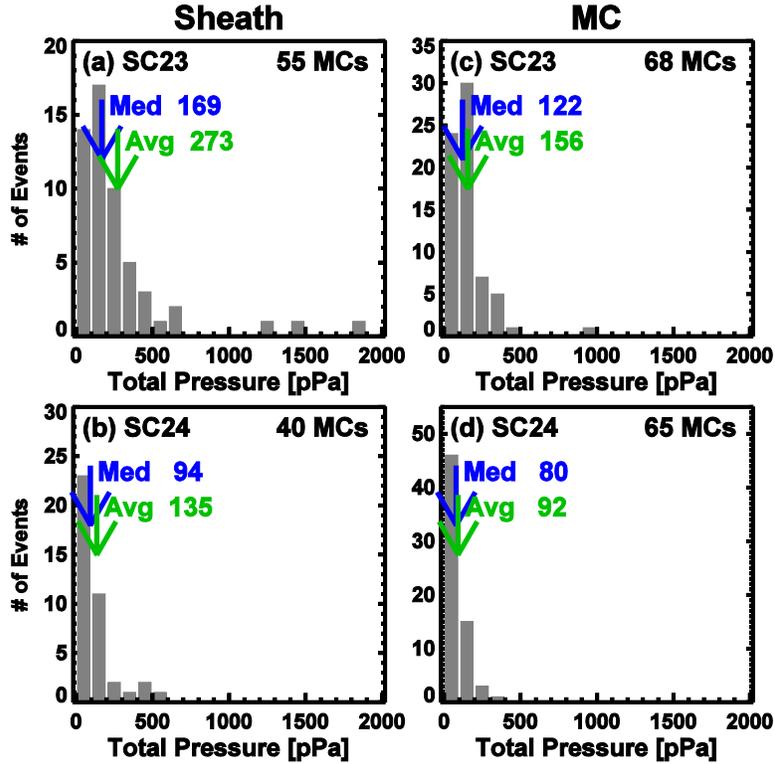

**Figure 14**. Comparison between sheath (a,b) and MC (c,d) total pressures (magnetic +plasma) for solar cycles 23 (top) and 24 (bottom). The mean and median values are noted on the plots.

A high-speed interplanetary structure with negative $B_z$ is known to be highly geoeffective. Gopalswamy et al. [2014a] reported that the lack of major geomagnetic storms in cycle 24 stems from the reduced magnitude of the $B_z$ component of the interplanetary magnetic field. In order to get further insight into the low *Dst* values in cycle 24, we have shown various scatter plots of the *Dst* index with $B_t$, $B_z$, $V$, and the product $VB_z$ in Fig. 13. As expected, the highest correlation was found between *Dst* and $VB_z$ in the cloud portion for both cycles 23 ($r = -0.77$) and 24 ($r = -0.86$). These correlations are highly significant because the critical value of the Pearson correlation coefficient for $P=0.05$ is 0.255 for MCs and 0.282 for sheaths. Figure 13 confirms that $VB_z$ is the primary factor determining the strength of a geomagnetic storm measured by the *Dst* index (Wu and Lepping 2002; Echer et al. 2005; Gopalswamy et al. 2008; Gopalswamy 2010). The *Dst*-$VB_z$ relationship has not changed much, suggesting that the efficiency of the process that generates geomagnetic storms has not changed. The narrow range of $VB_z$ in cycle 24 resulted in the small value of the maximum storm strength.

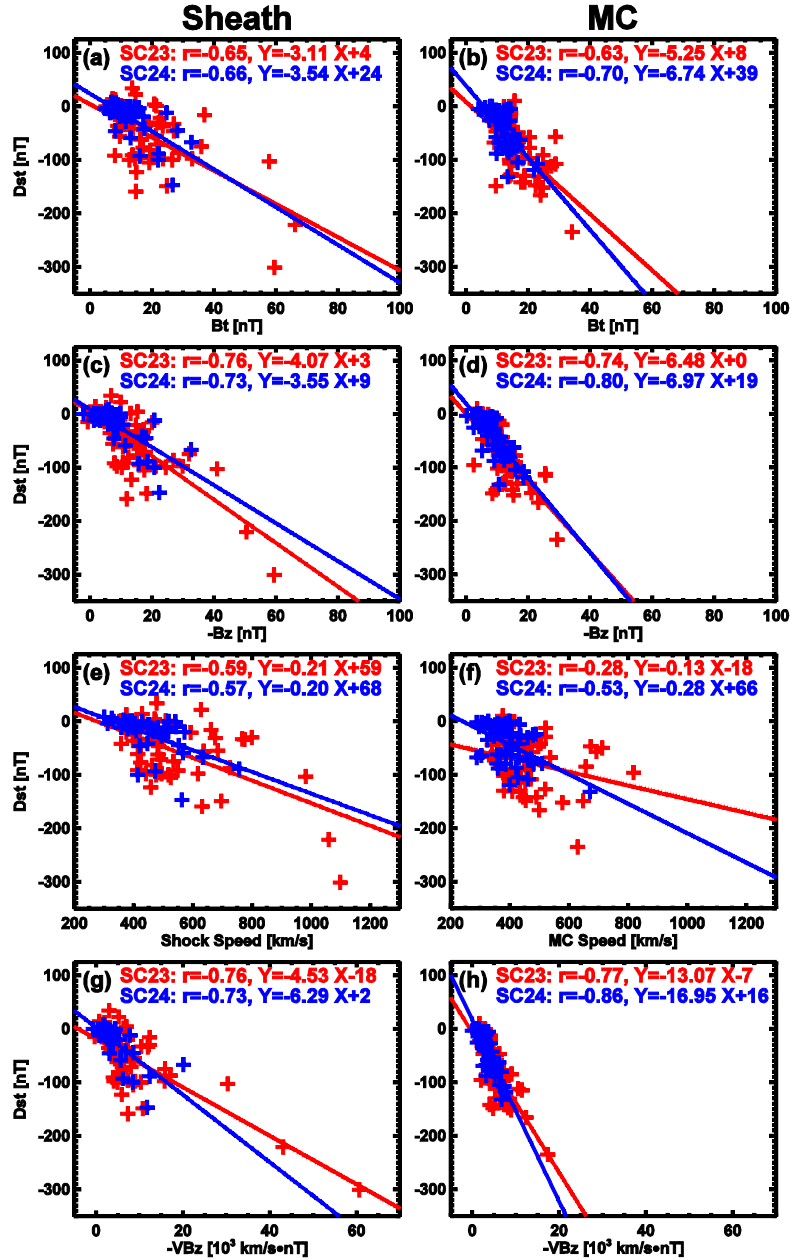

**Figure 13**. Scatter plots of *Dst* with $B_t$, $B_z$, $V$, and the product $VB_z$ for the sheath (left column) and cloud (right column). Events from cycle 23 and 24 are distinguished by red and blue symbols, respectively with the corresponding colors for the regression lines. In the right-side plots, the FN type MCs are not included because they are not geoeffective.

### 3.2.5 Total Pressure

The total pressure $P_t$ inside the magnetic clouds is an important quantity that determines how MCs expand [see e.g. Démoulin and Dasso 2009]. In particular, the difference in $P_t$ between the magnetic cloud and the ambient medium dictates the expansion rate since the gas pressure inside the clouds is generally much smaller than the magnetic pressure. Gopalswamy et al. [2014a] reported that the cycle-24 total pressure of the ambient medium at 1-AU was ~38% smaller than that in cycle 23. Therefore, it is important to see how the pressure inside MCs compares with the ambient pressure. We computed $P_t$ in sheaths and MCs using the magnetic field strength, proton density, and temperature measurements with the usual assumptions [Jian et al. 2006; Gopalswamy et al. 2014a]: $B^2_t/8\pi$ was added to the plasma pressure $N_p k T_p$ corrected for electron and alpha particle contributions. The resulting $P_t$ for the two cycles are shown in Fig. 14. The most probable values of the distributions are in lower bins in cycle 24. The total pressures in the sheath and clouds are significantly reduced in cycle 24 by ~ 50% in the sheath and 41% in the cloud portions. KS test resulted presented in Table 2 and 3 confirm that the $P_t$ difference in the two cycles is statistically significant.

### 3.3 Summary of Comparisons

From Tables 2 and 3 we see that most of the MC and sheath parameters are significantly different between the two cycles. The cycle-24 decrease in sheath density (15%) and duration (14%) are not statistically significant (as indicated by $D < D_c$). In the case of MCs, the proton density and MC duration do not differ significantly between the two cycles (see Table 2). In addition, the expansion factor and another quantity $\zeta$ (to be defined in the next section) related to the expansion factor also do not show statistically significant differences. The implications of these results are discussed in the following section.

### 4. Discussion

We compared the properties of MCs in cycle 23 and 24 and found several significant differences, all pointing to the weak solar activity in cycle 24. Some changes such as the MC magnetic content ($B_t$, $B_z$, $P_t$) are drastic, but some others did not show much change. In particular, the similar number of MCs and similar MC durations stand out. The similarity in MC numbers in the two cycles was also seen in the number of full halo CMEs observed in the two cycles [Gopalswamy et al. 2015a] because a majority of MCs are associated with halo CMEs [Gopalswamy et al. 2010b]. Therefore, the MC number in cycle 24 is consistent with the higher occurrence rate of halos relative to the sunspot number. One of the explanations for the higher abundance of halos in cycle 24 is the increased expansion of CMEs in the weak heliosphere of cycle 24 [Gopalswamy et al. 2015a]. The MC duration also did not drop significantly (see Table 2), although the size of the MCs dropped by ~22%. The MC duration $\Delta t \approx size/V$, so similar decline in the size and speed of MCs (see Table 2) must have resulted in a similar $\Delta t$ in the two cycles. The expansion factor is the ratio of the leading-edge speed to the central speed of MCs. Both of these speeds declined by the same extent, so the expansion factor remains unchanged. The average proton density in cycle-24 MCs decreased only slightly (~9%), which is not significant; the drop in the ambient density (~20%) between the two cycles is much larger [see e.g., Gopalswamy et al. 2014a, their fig.3]. Since the magnetic pressure in MCs is dominant, the differences in plasma properties seem to be very small between the solar cycles.

In two respects the behavior of CMEs near the Sun and MCs at 1 AU are different: (i) The CME speeds at the Sun are the same in cycles 23 and 24 [Gopalswamy et al. 2014a], whereas the cycle-24 MCs are significantly slower by ~15%. (ii) The CME widths are higher in cycle 24 near the Sun whereas the 1-AU MC sizes are smaller by 22%. The drop in MC speeds between the two cycles can be readily explained in terms of the increased drag force due to wider CMEs near the Sun and the slower solar wind in cycle 24 [McComas et al. 2013; Fig. 16].

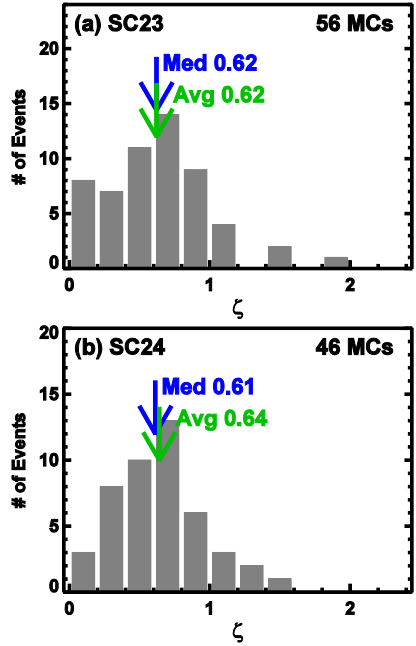

**Figure 15**. Histogram of the dimensionless expansion rate $\zeta$ for cycle 23 (top) and 24 (bottom). Only clouds with $V_{exp}>0$ are included.

## 4.1 MC Expansion Rate and Total Pressure

Updating the results of Gopalswamy et al. [2014a], we find that the average total pressure of the ambient medium at 1 AU remains lower in cycle 24 even though the solar activity has peaked and started declining. The average total pressures at 1 AU are 42.4 and 29.6 pPa, respectively in the first 73 months of cycles 23 and 24 indicating that the cycle-24 total pressure is reduced by ~30%. In section 3.6, we saw that the average total pressure $P_t$ inside MCs dropped by 34% from 139.8 pPa to 91.7 pPa in cycle 24, very similar to the drop in the ambient pressure. The $P_t$ ratio between MCs and the ambient medium remains roughly the same: 3.30 (cycle 23) and 3.10 (cycle 24).

Since this difference between the MC- and the ambient-medium pressures is responsible for the expansion of MCs, it is instructive to compare the dimensionless expansion rate defined as $\zeta = V_{exp}L/\Delta t V^2_c$, where $L$ is the heliocentric distance of the MC [Démoulin and Dasso 2009]. Recall that $\Delta t$ remained roughly the same, so for $L = 1$ AU, we expect $\zeta \sim V_{exp}/V^2_c$. Since we have all the numbers that make up $\zeta$, we computed it for each MC in the two cycles and the distribution is shown in Fig. 15. We see that the average expansion rates in the two cycles are identical, consistent with the identical cloud-to-ambient pressure ratios. Démoulin and Dasso [2009] had used the X-component of the MC speed, while we have used the total speed. For MCs heading toward Earth, we expect the difference to be insignificant. Since $V_{exp}$ is given by the difference ($\Delta P_t$) in total pressure between the ambient medium and the MC, we see that $\zeta \sim \Delta P_t/V^2_c$. The constancy of $\zeta$ can be understood from the fact that the reduction in $\Delta P_t$ in cycle 24 is compensated by the reduction in MC central speed ($V_c$).

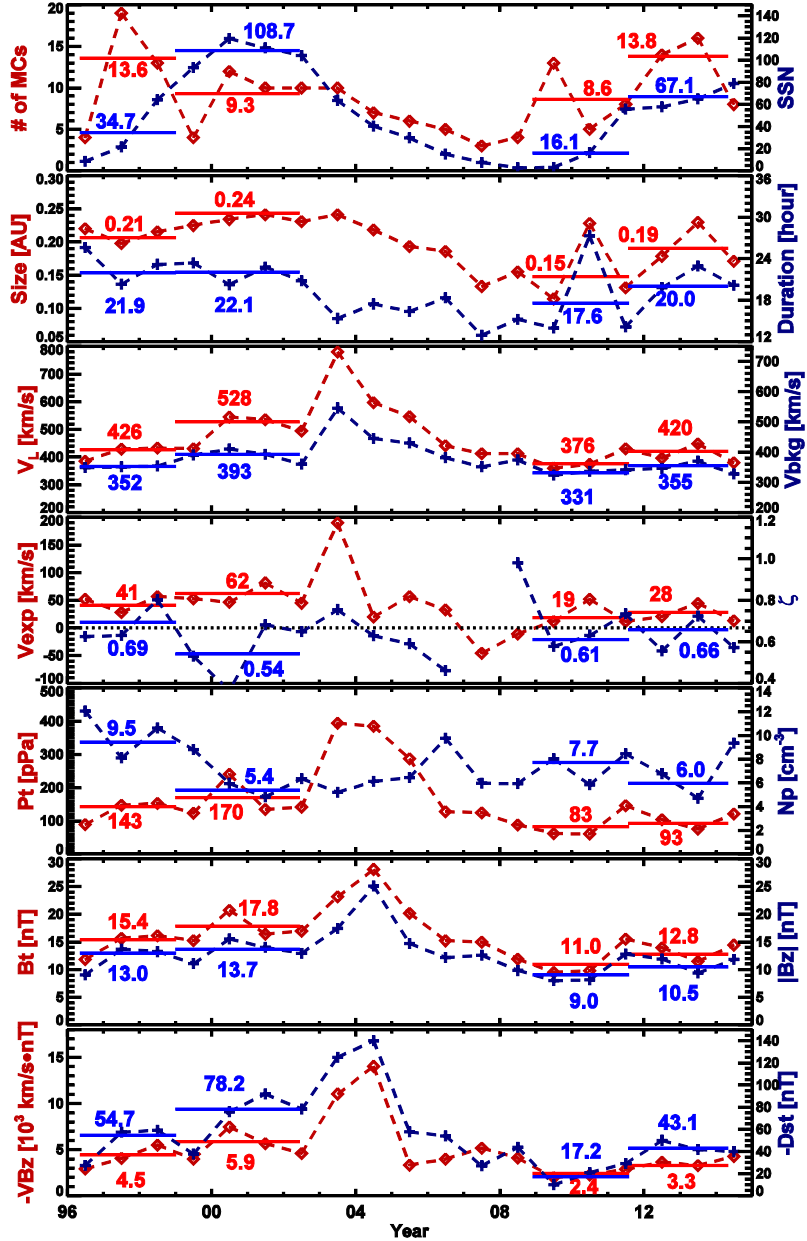

**Figure 16**. Solar-cycle variation of several parameters considered in this paper. The number of MCs and SSN are given in the top panel for reference. The red (blue) curves refer to the left (right) Y-axis. The first set of horizontal lines correspond to the rise and maximum phases of cycle 23, while the second set is for cycle 24. *Vbkg* is the background solar wind speed computed as the minimum value in a 12-h interval ahead of the shock (MC if not shock-driving). The large spike around 2003-2004 is due to Halloween 2003 events and other energetic events in 2004.

Since the 1-AU expansion rate $\zeta$ is nearly identical in the two cycles, we conclude that the anomalous expansion of CMEs must occur very close to the Sun before CMEs acquire constant widths in the coronagraph FOV. Gulisano et al. [2010] suggested that the size (*S*) of MCs at a distance *L* from the Sun depends on the expansion factor $\zeta$: $S = S_o (L/L_o)^\zeta$, where $S_o$ is the initial size of the flux rope at a distance $L_o$ from the Sun. It is also assumed that $\zeta$ does not depend on distance beyond $L_o$, which is the distance where an approximate pressure balance between the flux rope and the ambient medium is reached. The MC size ratio for the two cycles is $S_{24}/S_{23} = (S_{o24}/S_{o23})(L_{o23}/L_{o24})^\zeta$. Although we do not know the initial flux rope sizes, we can estimate the ratio based on the fact that cycle-24 CMEs were wider near the Sun for a given CME speed [Gopalswamy et al. 2014a]. CMEs that become MCs typically have a speed of ~1000 km s$^{-1}$ and for these CMEs, the cycle-24 width ($W_{24}$) is larger than that in cycle 23 ($W_{23}$) by 38%. The white-light CME width used by Gopalswamy et al. [2014a] is in the sky plane, providing the angular extent of the CME, which includes both the main body (flux rope) of the CME and the surrounding disturbance. CME width typically increases in the first few solar radii, and then stabilizes to a quasi-constant value. If the flux rope is in the sky plane, the flux-rope opening angle is likely to be proportional to the measured CME width ($W$). If we assume that the increase in CME width is the same as the increase in flux rope size, we can take $S_{o24}/S_{o23} = W_{24}/W_{23} = 1.38$. From Table 2, we have $S_{24}/S_{23} = 0.78$. Since $\zeta = 0.64$ (Fig. 15) we can determine $L_{o23}/L_{o24}$ as 0.5. This means the observed size difference between cycles 23 and 24 can be explained if the pressure balance is attained at a slightly larger distance from the Sun. It is known from cycle-23 observations that the CME mass and width steadily increase until ~5 solar radii and then level off [see e.g., Vourlidas et al. 2002; Gopalswamy 2004; Yashiro et al. 2004]. This is likely to be the distance where the pressure balance is attained. This means $L_{o24}$ ~10 solar radii. This is consistent with the increased near-Sun CME expansion in cycle 24, so CMEs travel larger distances before pressure balance is attained. It is possible to verify this by selecting limb CMEs with similar neutral-line orientation in the two cycles and measure the width variation with height. Results of such an investigation will be reported elsewhere.

**4.2 Solar Cycle Variation**

In the above discussion we compared MC properties between cycles 23 and 24 using 73-month averages. It is well-known that CME speeds show variation within various phases of a solar cycle. For example, CME speeds and occurrence rates are higher during solar maximum [e.g., Gopalswamy 2004]. The ratio of the number of MCs to the number of non-cloud ICMEs also show a solar cycle variation, with a smaller fraction of MCs in the maximum phase [Riley et al. 2006]. The source location of CMEs associated with MCs follows the sunspot butterfly diagram [Gopalswamy et al. 2008]. In this subsection, we consider solar cycle variation of most of the MC parameters considered in Table 2, over a period from May 1996 to December 2014. The study period covers the rise, maximum, and decay phases of cycle 23, followed by the rise and maximum phases of cycle 24 (with only a small portion of the decay phase). We have data spanning almost two solar cycles, so we can study intra-cycle as well as inter-cycle variations. As noted before, the length of the phases in the two cycles differ slightly. The cycle-23 rise and maximum phases are 32 months and 41 months long, respectively. In cycle 24, the lengths are 32 months (rise) and 45 months (maximum). While the averages are shown only for the individual phases, the yearly averages are shown for the entire cycle 23 (i.e., including the declining phase) and the cycle 24 to the end of 2014.

Figure 16 shows the annual and phase averages of a dozen quantities associated with MCs along with the annual number of MCs and SSN for reference. The discordance between SSN and MC number is clear both at inter-cycle and intra-cycle levels. The number of MCs is higher in the rise phase compared to the maximum phase in cycle 23, while the opposite is true for cycle 24. SSN jumps by a factor of ~4 from rise to maximum phases in cycle 24, compared to a factor of ~3 in cycle 23. The 73-month averages discussed earlier are mostly due to the difference in the maximum phases of the two cycles; cycle-24 rise phase is much weaker than the cycle-23 one (by 54%). Many of these quantities show solar cycle variation, confirming Dasso et al. [2012] and Lepping et al. [2015] when similar parameters are considered. The parameters ($V_L$, $B_t$, $P_t$, $VB_z$, $B_z$) that are directly linked to the solar source show clear solar cycle variation, except for the large modulation caused by periods of extreme activity in the decay phase of cycle 23, viz., October-November 2003 Halloween eruptions [Gopalswamy et al. 2005a] and November 2004 eruptions [Gopalswamy et al. 2006; Echer et al. 2010]. The time profiles of these quantities closely track each other. The *Dst* index closely follows the variation in $VB_z$ because the two quantities are highly correlated. The MC size varies smoothly over the whole of cycle 23, but shows marked fluctuations in cycle 24, in agreement with Lepping et al. [2015]. The expansion speed also shows a weak solar cycle variation, with a large spike during the Halloween 2003 period. The speed of the undisturbed solar wind $V_{bkg}$ in the upstream of MCs (or shocks) is a proxy to the background solar wind speed and is also smaller in cycle 24 but only by a small amount (<10%).

Among the parameters plotted in Fig. 16, three of them (MC duration $\Delta t$, expansion rate $\zeta$, and MC proton density $N_p$) have been found to vary insignificantly between the first 73 months of cycles 23 and 24 (see Table 2). The annual averages of $\Delta t$ roughly follow the MC size, but the range of variability is small, consistent with the results in Table 2. The annual averages of $\zeta$ also vary very little, confined between 0.58 and 0.69. There was a gap in 2007 because the clouds were not expanding, so we did not compute the expansion factors. There is no solar cycle variation in the annual averages of $N_p$, which appears almost flat over the entire period plotted in Fig. 16.

The horizontal bars in Fig. 16 mark the solar cycle phases and the number near them is the average over the phase. The averages are shown only for the rise and maximum phases of cycles 23 and 24. It is clear that all parameters show an increase from the rise to the maximum phase in both cycles, except for the three non-varying ones: $\Delta t$, $\zeta$, and $N_p$. In cycle 23, $\Delta t$ remains constant between rise and maximum phases, but there is an increase of 14% from the rise to the maximum phase in cycle 24. There is no specific trend in the variation of $\zeta$: while it declines from the rise to the maximum phase in cycle 23, the opposite happens in cycle 24. Only $N_p$ shows consistent decline from the rise to the maximum phases in both cycles. When we compare parameters between the same phases in the two cycles, we see that the cycle 23 values are greater than those in cycle 24 for almost all of them. One of the striking variation between phases is the increase in the *Dst* index averaged over the rise and maximum phases in cycle 24: the average value more than doubled from -17.2 nT to - 43.1 nT. While the average value in the maximum phase is barely at the storm level, the rise phase value is well below the storm level (no storms). In some cases, the average values during the maximum phase of cycle 24 are below the average values of the minimum phase of cycle 23 (e.g., $B_t$ and $VB_z$), confirming the weakness of cycle 24. In cycle 23, the average *Dst* following MCs is at the storm level for both rise and maximum phases. As in cycle 24, more intense storms are found in the maximum phase, in agreement with the finding of Kilpua et al. [2015], who reported that more intense storms tend to occur during the maximum phases of solar cycles.

The drastic reduction in *Dst* in cycle-24 MCs is consistent with the corresponding reduction in the magnitude of $VB_z$. The rise phase of cycle 24 has no geoeffective CMEs at all because the average *Dst* index is only 17.2 nT (recall that the Dst index following an MC needs to be < -30 nT to be geoeffective). On the other hand there are many geoeffective MCs in the rise phase of cycle 23, with an average *Dst* of 54.7 nT. Comparing the maximum phases in the two cycles, we see that the drop in the average $VB_z$ (44%) and the associated *Dst* index (45%) are very similar, strongly confirming the suggestion by Gopalswamy et al. [2014a] based on major storms (*Dst* ≤ -100 nT). Here we have confirmed it taking the reverse path of starting with the MC properties. We suggest that the dilution in the magnetic content of the MCs occurs close to the Sun, which persists as the CME evolves through the interplanetary medium into MCs observed in situ. We appreciate that there are a lot more ICMEs than MCs in both cycles. But we chose to study only MCs because their magnetic structure is known and their nose region crosses Earth. Furthermore, the analysis may also apply to non-cloud ICMEs because all ICMEs seem to contain flux ropes but not observed as such due to observational limitations [Gopalswamy et al. 2013; Mäkelä et al. 2013, Xie et al. 2013; Kim et al. 2013].

### 4.3 Extreme Events

It must be pointed out that extreme events do occur at any phase of a given cycle or the cycle amplitude. In the well-known backside event of 2012 July 23, the MC arrived in ~19h at the Solar Terrestrial Relations Observatory's Ahead spacecraft with a Bz of ~ -52 nT and a speed of ~1500 km s$^{-1}$ [Russell et al. 2013; Baker et al. 2013; Mewaldt et al. 2013; Liu et al. 2014; Gopalswamy et al. 2014b]. If the CME were directed toward Earth, it is estimated that a Carrington-size geomagnetic storm would have occurred (for a $VB_z$ of -7.8×10$^4$ km s$^{-1}$.nT, the *Dst* estimate was ~ -800 nT using the empirical relation (*Dst* = -0.01$VB_z$ - 25 nT) from Gopalswamy et al. [2008]). If we use the relationship in Fig. 13h derived for cycle 24 (viz., *Dst* = -0.017 $VB_z$ +16 nT) the expected *Dst* is ~ -1300 nT. If the $VB_z$ value is modified to account for the average 40% reduction in cycle 24 (see Table 2), the new $VB_z$ is -7.8×10$^4$ km s$^{-1}$.nT/0.6 or -1.3×10$^5$ km s$^{-1}$.nT. This results in a *Dst* of -2200 nT, which is close to the largest storm magnitude possible given the size of Earth's dipole [Vasyliunas 2011]. In other words, our estimate indicates that the weak heliosphere is likely to have mitigated the storm intensity.

The largest storm of cycle 24, which occurred on 2015 March 17 was due to the 2015 March 15 CME that resulted in an MC with a $B_z$ of -25 nT that arrived at Earth with a speed of ~600 km s$^{-1}$. The MC caused a geomagnetic storm that had a minimum *Dst* of -223 nT (real time data from WDC Kyoto). The regression equation for cycle 24 in Fig. 13h (*Dst* = -0.017$VB_z$ + 16 nT) yields a *Dst* of -239 nT, not too different from the observed value. In other words, the dependence of *Dst* on $VB_z$ remains the same in this cycle, but the $VB_z$ is diminished. Had the same MC occurred in cycle 23, it would have resulted in a $VB_z$ of -2.5×10$^4$ km s$^{-1}$.nT/0.6 or -2.5×10$^4$ km s$^{-1}$.nT and hence a *Dst* of -408 nT. Interestingly, the largest storm of cycle 23 had a *Dst* of -403 nT [Gopalswamy et al. 2005b].

### 5. Summary and Conclusions

The main purpose of this study was to compare the properties of magnetic clouds between cycles 23 and 24 in order to understand their difference in their geoeffectiveness. In particular, we seek to understand the mild space weather in cycle 24 as indicated by the low *Dst* index. We found some significant differences between MCs of cycles 23 and 24 (and between the sheaths, if the MCs were shock-driving). There are also variations within each cycle (intra-cycle variations) between the rise and maximum phases. However, five quantities did not change significantly between the two cycles: (1) the number of magnetic clouds over the first 73 months in each cycle (similar to the number of front-side halo CMEs), (2) the sheath and cloud proton densities, (3) the sheath and cloud durations, (4) the expansion factor of the MCs, and (5) the dimensionless expansion rate of MCs. The main conclusions are listed below.

(i) Magnetic clouds show both intra-cycle and inter-cycle variations in properties, but there is a clear discordance between the number of MCs and the sunspot number. There are more MCs per SSN in cycle 24, similar to halo CMEs.

(ii) The number of shock-driving MCs is smaller in cycle 24 (62% vs. 81% in cycle 23), although the CME speeds near the Sun have been reported to be nearly the same in the two cycles [Gopalswamy et al. 2014a].

(iii) Even though cycle 24 is weak as indicated by the sunspot number, the cloud types show normal progression: the bipolar MCs (north-south and south-north types) follow the 22-year cycle between polarity reversals at solar poles.

(iv) The parameters ($V_L$, $B_t$, $P_t$, $VB_z$, $B_z$) are directly linked to the solar source and show clear solar cycle variation, except for the large modulation caused by periods of extreme activity in the decay phase of cycle 23. The solar-cycle variation is in the sense that these parameters have higher values in the solar maximum phase compared to the rise phase.

(v) The Dst index associated with MCs shows a solar cycle variation very similar to that of $VB_z$. There is a clear tendency for larger storms to occur in the maximum phase.

(vi) The expansion speed of MCs in cycle 24 is smaller than that in cycle 23, consistent with the diminished MC-to-ambient total pressure difference ($\Delta P_t$) in cycle 24. The expansion speed remains a small fraction of the MC leading edge speed.

(vii) The dimensionless expansion rate $\zeta$ remains constant between the two cycles because the reduction in MC-to-ambient total pressure difference ($\Delta P_t$) is compensated for by the reduction in MC central speed.

(viii) The size of the cycle-24 MCs at 1 AU are significantly smaller, contrary to the white-light CME sizes near the Sun. Based on the observed sizes at 1 AU in cycles 23 and 24, we suggest that the increased CME expansion in cycle 24 near the Sun resulted in a larger pressure-balance distance: ~10 solar radii compared to ~5 solar radii in cycle 23.

(ix) The peak magnetic field strength ($B_t$) in the sheath and cloud portions are reasonably correlated with the corresponding speeds in both cycles. The correlation coefficients are: 0.80 and 0.49 for sheath and cloud portions in cycle 23; the corresponding numbers are 0.63 and 0.36 for cycle 24.

(x) The average *Dst* index in the sheath and cloud portions in cycle 24 was -33 nT and -23 nT, compared to -66 nT and -55 nT, respectively in cycle 23. The *Dst* index associated with MCs in the rise and maximum phases of cycle 24 show that the rise-phase MCs are not geoeffective at all.

(xi) The Dst - $VB_z$ correlation is the highest in both cycles. The correlation coefficients are: 0.76 and 0.77 for the sheath and cloud portions in cycle 23; the corresponding numbers are 0.73 and 0.86 for cycle 24.

(xii) The empirical relationship between the *Dst* index and the $VB_z$ in MCs continues to hold, suggesting that the efficiency of the process causing geomagnetic storms has not changed significantly between the cycles. However, the drastic reduction in $VB_z$ (by 40%) seems to be primarily responsible for a similar reduction in the storm strength in cycle 24. The reduction in the MC speed and $B_z$ has contributed to the reduction in $VB_z$.

(xiii) The CME expansion near the Sun is likely to have caused the weak magnetic content in the flux ropes (magnetic pressure, field strength, and the out-of-the-ecliptic component) as observed at 1 AU, but the dilution seems to have occurred close to the Sun.

**Acknowledgments:**

We thank T. Nieves-Chinchilla (http://wind.nasa.gov/index_WI_ICME_list.htm) and I. G. Richardson (http://www.srl.caltech.edu/ACE/ASC/DATA/level3/icmetable2.htm) for making ICME lists available online. The parameters used in this paper were derived from the OMNI data available online at NASA Goddard Space Flight Center (http://omniweb.gsfc.nasa.gov). We thank R. Skoug for providing *ACE* data for three events (Bastille Day 2000, Halloween 2003). The *Dst* index was obtained from the World Data Center, Kyoto, Japan ((http://wdc.kugi.kyoto-u.ac.jp/dstdir/index.html). The work of NG, SY, SA was supported by NASA/LWS program. PM was partially supported by NSF grant AGS-1358274 and NASA grant NNX15AB77G. HX was partially supported by NASA grant NNX15AB70G.